\documentstyle[11pt,aaspp4,epsfig]{article}

\begin{document}

\def\simg{\mathrel{%
      \rlap{\raise 0.511ex \hbox{$>$}}{\lower 0.511ex \hbox{$\sim$}}}}
\def\siml{\mathrel{%
      \rlap{\raise 0.511ex \hbox{$<$}}{\lower 0.511ex \hbox{$\sim$}}}}
\def\Mesz{M\'esz\'aros~}
\def\beq{\begin{equation}}
\def\enq{\end{equation}}
\def\bea{\begin{eqnarray}}
\def\ena{\end{eqnarray}}
\def\bec{\begin{center}}
\def\enc{\end{center}}
\def\etal{{\it et al.~}}
\def\ro{r_o}
\def\ergsi{\hbox{erg s$^{-1}$}}
\def\cmcui{{\rm cm}$^{-3}$}
\def\msun{M_\odot}
\def\eps{\epsilon}
\def\varep{\varepsilon}
\def\refe{\reference}
\def\cm{{\rm cm}}
\def\cmsqi{{\rm cm}^{-2}}
\def\cmcui{{\rm cm}^{-3}}
\def\gcmcui{{\rm g~cm}^{-3}}
\def\si{{\rm s}^{-1}}
\def\s{\rm s}

\def\L52{L_{52}}
\def\Eint{E_{int}}
\def\Epm{E_{\pm}}
\def\Eiso{E_{iso}}
\def\st{\sigma_T}
\def\rpm{r_\pm}
\def\E54{E_{54}}
\def\Egiso{E_{\gamma,iso}}
\def\Egam{E_{\gamma}}
\def\Lgiso{L_{\gamma,iso}}
\def\Lgam{L_{\gamma}}
\def\Lsh{L_{sh}}
\def\Ex{E_{x}}
\def\epse3{\epsilon_{e,1/3}}
\def\epsb3{\epsilon_{B,1/3}}
\def\epsi4{\epsilon_{i,1/4}}
\def\GM3{\Gamma_{M,3}}
\def\Gm2{\Gamma_{m,2}}
\def\Gc25{\Gamma_{c,2.5}}
\def\tvpm{t_{v\pm}}
\def\tvm3{t_{v,-3}}
\def\tv3{t_{v,-3}}
\def\etast{\eta_\ast}
\def\etat{\eta_t}
\def\ro7{r_{o,7}}

\title{X-ray Rich GRB, Photospheres and Variability}

\author{P. \Mesz$^{1,2}$, E. Ramirez-Ruiz$^3$, M.~J. Rees$^3$ \& B.
Zhang$^1$}

\noindent
$^1${Dpt. of Astronomy \& Astrophysics, Pennsylvania State University,
University Park, PA 16803}\\
\smallskip\noindent
$^2${Dpt. of Physics, Pennsylvania State University,
University Park, PA 16803}\\
\smallskip\noindent
$^3${Institute of Astronomy, University of Cambridge, Madingley Road,
Cambridge CB3 0HA, U.K.}

\bec{ApJ, submitted: 5/8/02; accepted: 6/20/02}\enc

\begin{abstract}

We investigate the relationship between the quasi-thermal baryon-related
photosphere in relativistic outflows, and the internal shocks arising
outside them, which out to a limiting radius may be able to create enough
pairs to extend the optically thick region. Variable gamma-ray light
curves are likely to arise outside this limiting pair-forming shock radius, 
while X-ray excess bursts may arise from shocks occurring below it; a 
possible relation to X-ray flashes is discussed. This model leads to a 
simple physical interpretation of the observational gamma-ray
variability-luminosity relation.

\end{abstract}


\section{Introduction}

Gamma-ray burst (GRB) light curves at $\gamma$-ray energies are often
highly variable, and generally this is attributed to internal shocks
occurring at some radius $\simg 10^{12}-10^{14}$ cm from the center
of a relativistic outflow produced by a violent collapse or merger,
beyond the ``photospheric" radius at which the flow becomes optically
thin to scattering by electrons associated with the baryons entrained.

This photosphere is a source of soft thermalized radiation, which may be
observationally detectable in some GRB spectra, and may also result in
inverse Compton cooling of the non-thermal electrons accelerated in the
shocks occurring outside it, thereby enhancing a hard GeV non-thermal
component at the expense of the usual MeV synchrotron component.

At small enough radii, however, the shocks can create enough pairs to
re-establish a second photosphere caused by the pairs, and it is only
beyond a limiting radius that the shocks remain optically thin to pairs.
The most favorable region for shocks producing highly variable gamma-ray
light curves is above this radius, while shocks occurring below it would
lead to a second source of less variable radiation (Kobayashi, Ryde \&
MacFadyen 2002; see also Ramirez-Ruiz \& Lloyd-Ronning 2002; and Spada,
Panaitescu \& \Mesz 2000),
which is also X-ray rich.  Bursts dominated by either the baryonic
photosphere or by pair-producing shocks may be identified with X-ray
excess bursts (Preece \etal 1996), while the latter resemble at least
the harder examples of the proposed X-ray flash (Heise \etal 2001)
sub-class of GRB.

The existence of a limiting pair-forming shock radius provides also a
scenario which combines recent work on the interpretation of afterglow
light-curve breaks in terms of a jet opening angle (Frail \etal 2001;
Panaitescu \& Kumar 2001; Piran \etal 2002) or a universal jet shape
(Rossi, Lazzati \& Rees 2002; Zhang \& \Mesz 2002; Salmonson \& Galama
2002), and
work on the gamma-ray variability-luminosity relationship (Fenimore \&
Ramirez-Ruiz 2001; Reichart \etal 2001). Identifying the pair shock
radius as an approximate boundary above which shocks lead to more
strongly variable gamma-ray light curves and below which shocks result
in smoother X-ray rich light curves, a phenomenological jet model leads
to a simple physical explanation of the quantitative form of the
variability-luminosity relationship.

\section{Baryonic Photospheres}
\label{sec:barphot}

Consider a relativistic wind outflow where the bulk Lorentz factor
has a mean dimensionless  entropy $\eta=L_o/{\dot M} c^2 =10^2\eta_2$
which varies ($\Delta\eta \sim \eta$) on timescales $t_v$ ranging from
a minimum dynamical timescale up to the maximum burst (wind) duration
$t_w$,
$10^{-3}~s \leq t_v \siml t_w$. The flow starts from a minimum radius
$r_o=c t_{v,min} =10^7 \ro7$ cm, and the Lorentz factor accelerates
as $\Gamma \propto r$ up to a coasting (or saturation) radius $r_c \sim
r_o\Gamma_f$, beyond which it coasts as $\Gamma = \Gamma_f $. For a simple
wind, neglecting finite shell effects, $\Gamma_f=\min[\eta,\etast]$ where
the value $\eta_\ast$ is a critical value of the dimensionless entropy
given
by (\Mesz \& Rees 2000)
\beq
\eta_\ast \simeq \ell_{p,o}^{1/4} = ({L_o\st/4\pi m_p c^3 r_o})^{1/4}
\simeq
10^3 (\L52 \ro7^{-1})^{1/4}~.
\label{eq:etast}
\enq
Here $\ell_{p,o}$ is analogous to the definition of the compactness
parameter but using the proton instead of the electron mass.
The coasting $\Gamma_f$ values follow from the criterion that the proton
drag
time must be longer than the expansion time for protons to start to coast.
Below the baryonic photosphere protons are naturally coupled to radiation,
but in the optically thin region above the photosphere, if this occurs in
the accelerating regime, the protons can  still coupled to radiation
and continue to accelerate out to a radius beyond the photosphere.
The comoving density in the (continuous) wind regime is
$n'=(L_o / 4 \pi r^2 m_p c^3 \eta\Gamma)$,  and using the above
behavior of $\Gamma$ below and above the coasting radius, as well as
the definition of the Thompson optical depth in a continuous wind
$\tau_T\simeq n'\st(r/\Gamma)$ we find that the baryonic photosphere where
$\tau_T=1$ in the wind (w) regime, due to electrons associated with
baryons, is
\beq
 {r_{ph,w}\over r_o}=\cases{
   \eta_\ast^{4/3} \eta^{-1/3} ~&~ for $r<r_o\eta$; \cr
   \eta_\ast^4 \eta^{-3}       ~&~ for $r>r_o\eta$,\cr}
\label{eq:rphw}
\enq
where for a wind $r_c\equiv r_o \min[\eta,\etast]$ is the coasting radius
beyond which $\Gamma=\Gamma_f=$ constant.

The accelerating and coasting behavior is followed on average also if the
outflow consists of shells of duration $t_v=\Delta_o/c \geq r_o$,
separated by intervals which could similarly be of order $\sim t_v$ (or a
superposition of several such frequencies), leading to an oscillatory
modulation of the linear and coasting behavior. Aside from such
modulation,
in the optically thick regime the Lorentz factor can never exceed $\Gamma
\leq \eta$. However taking into account the finite shell structure, in the
optically thin regime the coasting $\Gamma_f$ will differ for some $\eta$
from the values $\min[\eta,\etast]$ discussed in the wind problem. Taking
for simplicity shells resulting from the minimum variation timescale
$\Delta_o= c t_{v,min}= r_o$, the evolving comoving  width of the shell
is $\Delta'=[r, ~r_o\eta,~ \eta^{-1} r ]$ and the comoving volume of the
shell is
$V'=4\pi r^2\Delta'=[4\pi r^3,~4\pi\eta r_o r^2,~4\pi \eta^{-1} r^3]$
when $r$ is $[<r_o\eta, ~>r_o\eta,~ >r_o\eta^2]$.
The comoving particle density $n'=(L t_v/ \eta m_p c^2 V')$ and the
Thomson depth $\tau_T=n'\st\Delta'=1$ define a baryonic photosphere
in the discrete shell (ds) regime
\beq
{r_{ph,ds}\over r_o}= \eta_\ast^2 \eta^{-1/2}~~~~
       \hbox{for both~} r<r_o\eta ~\hbox{and~} r>r_o\eta~~.
\label{eq:rphsh}
\enq
For a wind made up of shells of approximate duration $t_v$ ejected at
intervals of order $t_v$, at high $\eta$ the shells move fast enough that
a photon arising in one shell never crosses more than that one shell. At
lower $\eta$, however, a light ray can cross many shells before escaping,
and the appropriate expression for the photosphere approximates that of
the
wind equation (\ref{eq:rphw}). The criterion for the latter to be valid is
that $r \siml \Delta_o\eta^2 = r_o\eta^2$, and the transition occurs at
$\eta=\eta_t$ given by
\beq
\eta_t=\eta_\ast^{4/5}= 2.5\times 10^2 (\L52 \ro7^{-1})^{1/5}
\label{eq:etat}
\enq
Thus one has for the baryonic photosphere
\beq
{r_{ph}\over r_o}=\cases{
\etast^4\eta^{-3}  =\eta_t^5\eta^{-3} ~&~ for~ $\eta \leq \eta_t$;\cr
\etast^2\eta^{-1/2} =\eta_t^{5/2}\eta^{-1/2}~&~for~$\eta >\eta_t$, \cr}
\label{eq:rph-eta}
\enq
where the first (wind regime) occurs only in the coasting regime,
while the second (shell regime) applies partly in the accelerating and
partly in the coasting regimes.
These regimes differ from those in \Mesz \& Rees (2000) by having
a break at $\eta_t=\eta_\ast^{4/5}$ instead of $\eta_\ast$, and by having
a slope -1/2 above $\eta_t$ instead of -1/3 (due to the shell regime,
neglected in our previous paper).
The photosphere is in the coasting wind regime for $\eta \leq \etat$,
in the coasting shell regime for $\etat\leq\eta\leq \etast^{4/3}$, and in
the accelerating shell regime for $\eta\geq\etast^{4/3}$ (see Fig. 1).
When the photosphere is above the coasting radius, the final Lorentz factor
is just $\Gamma_f \sim \eta$. When the photosphere is below the coasting
radius, the baryons continue to be dragged by the radiation above the
photosphere until $t'_{drag} \sim (m_p c^2/c \sigma_T u'_\gamma)$ exceeds
$t'_{exp}\sim r/c\Gamma$, out to a radius $r_{drag}/r_o \leq \etast^2$
where $\Gamma \sim \min[\etast^2\eta^{-1/2},\etast]$. The final Lorentz
factor is thus $\Gamma_f= [\eta, \etast^2\eta^{-1/2},\etast]$ for values of
$[\eta <\etast^{4/3}, ~\etast^{4/3}<\eta <\etast^2, ~\eta >\etast^2]$.
This results in shock radii (Figure 1) which are divided into three regimes
(instead of the two in \Mesz \& Rees 2000, where in the wind regime the
values $\etat$ and $\etast^{4/3}$ were collapsed into a single $\etast$).


In units of the initial total luminosity $L_o$ and initial temperature
at $r_o$, $\Theta_o=kT_o/m_ec^2 \simeq 2 \L52^{1/4}\ro7^{-1/2}$ (i.e.
$T_o\sim 1~\L52^{1/4}\ro7^{-1/2}$ MeV), the lab-frame baryonic
photospheric
luminosity $L_{ph}$ and dimensionless temperature $\Theta_{ph}$ behave as
\beq
{L_{ph}\over L_o}={\Theta_{ph}\over\Theta_o}= \cases{
  (r_{ph}/ r_c)^{-2/3}= (\eta/\etast)^{8/3}=
       \etat^{-2/3}(\eta/\etat)^{8/3} ~&~for~$\eta<\etat$;\cr
  (r_{ph}/ r_c)^{-2/3}= \etast^{-1/3}(\eta/\etast)=
       \etat^{-2/3}(\eta/\etat)       ~&~for~$\etat
<\eta<\etast^{4/3}$;\cr
   1                                  ~&~for~$\eta>\etast^{4/3}$.\cr }
\label{eq:Lph-bar}
\enq
Thus $(L_{ph}/L_o)=(\Theta_{ph}/\Theta_o)= \etast^{-8/15}=
2.5\times 10^{-2}$ ($T_{ph} \sim 25/(1+z)$ keV in the observer frame)
for $\eta=\etat=\etast^{4/5} \sim 250$;
$(L_{ph}/L_o)=(\Theta_{ph}/\Theta_o)=\etast^{-1/3}= 10^{-1}$ for
$\eta= \etast=10^3$; and $(L_{ph}/L_o)=(\Theta_{ph}/\Theta_o)=1$
at $\eta=\etast^{4/3}\sim 10^4$ (where the photosphere occurs
at the coasting radius).

The internal shocks, which occur in the coasting regime at radii
$r/r_o=(\Delta_o/r_o)\eta^2 \geq \eta^2$, produce a shock photon
luminosity
\beq
\Lsh = \eps_e \eps_i L_o\sim 10^{-1}\epse3\epsi4 L_o~,
\label{eq:Lsh}
\enq
where the shock efficiency $\eps_{sh,-1}= 10^{-1}\epse3\epsi4$ is a
bolometric radiative efficiency when the cooling timescale is shorter
than the dynamical time. Similarly the magnetic luminosity (if the
turbulent field energy $\eps_B=(1/3)\epsb3$ is in equipartition with
that of randomized protons and electrons) is
$L_B = \eps_B \eps_i L_o\sim 10^{-1}\epsb3\epsi4 L_o \simeq \Lsh$.
Thus, $L_{ph} \ll \Lsh \sim L_B$ for $\eta<\etat\sim 250$;
$L_{ph} < \Lsh \sim L_B$ for $\etat <\eta < \etast \sim 10^3$; and
$L_{ph}> \Lsh \sim L_B$ for $\etast <\eta <\etast^{4/3} \sim 10^4$.
This means that for $\eta >\etast \sim 10^3$ the baryonic photospheric
component dominates the non-thermal internal shock component in a
bolometric sense. This will lead to inverse-Compton cooling of the
non-thermal electrons accelerated in the shocks, causing a weakening
and softening of the nonthermal synchrotron spectrum of the shock, at
the expense of a hard ($\simg$ GeV) inverse Compton component, while
most of the energy will be in a thermal X-ray component.

The BATSE $\gamma$-ray luminosity is broad-band in nature, and can
be written as $L_\gamma \sim (1/5)(1-\eps_{IC})\Lsh \siml (1/5)\Lsh$
where $\eps_{IC}$ is the IC efficiency, with a peak synchrotron frequency
depending on the comoving magnetic field value $B'$. For low values of
$\eta$, shocks occur closer in, leading to higher $B'$ and harder
synchrotron peaks.  For $\eta \sim \etat$ the baryonic thermal X-ray
photosphere may be responsible for the X-ray excess BATSE bursts (Preece
\etal 1996).  For lower $\eta$ the photospheric thermal peak is even
softer,
while the shocks occur closer in and produce harder synchrotron peaks
approaching the upper, less sensitive end of the BATSE band, which could
lead to an apparent dominance of the soft X-ray thermal photospheric peak.

For higher values $\etat \siml \eta \siml \etast$ the thermal peak tends
to blend with the synchrotron peak, resembling the canonical non-thermal
GRB spectrum, while for $\eta \simg \etast$ a hard ($\simg $ MeV) thermal
component would be predicted to dominate.

\section{Shocks Above the Pair-Radius and Variable $\gamma$-ray
lightcurves}
\label{sec:pairphot}

When shells of mass $(m/2)$ with Lorentz factors $\Gamma_1$ and $\Gamma_2$
collide, the mechanical efficiency for conversion of kinetic energy
$mc^2(\Gamma_1+\Gamma_2)$ into internal energy is $\eps_i=
(\Gamma_1 + \Gamma_2 -2\sqrt{\Gamma_1 \Gamma_2})/ (\Gamma_1+\Gamma_2)$,
where as before we parameterize $\eps_i =(1/4)\epsi4$. If a total of $2N$
shells are ejected which collide, and the total isotropic equivalent
kinetic energy of outflow is $\Eiso$, the corresponding internal energy
produced in the merger of two shells is $\Eint \sim  \eps_i N^{-1}
\Eiso$. Of that, a fraction $\eps_e$ is given to electrons, and for a
high radiation efficiency in the MeV range and a high compactness
parameter (i.e. high efficiency of pair formation) a fraction of
order 1/3 of the radiated energy could be converted into pairs,
and the energy in pairs in the merged shell is
\beq
\Epm \sim \eps_e\eps_i (1/3N) \Eiso \sim 10^{50.5}\epse3\epsi4
N_2^{-1}\E54~{\rm erg}.
\label{eq:Epm}
\enq
Assuming that $\Gamma$ in in the range
$[\Gamma_m,\Gamma_M]$, with $\Gamma_m < \Gamma_M$, the observed
radiation comes mainly from collisions involving shells at the
extremes of this range and is maximized for $\Gamma_m \ll \Gamma_M$.
Such merged shells move with a center of mass Lorentz factor
$\Gamma_c\sim\sqrt{\Gamma_m \Gamma_M}$.  For shells of initial
lab-frame widths $\Delta_o \sim r_o$, for radii above the shock
radius $r_{sh}= r_o \eta^2$  (which is also the ``expansion radius"
above which the comoving width $\propto r$ and the comoving volume
$\propto r^3$) the energy radiated in the shocks can be enough to create
pairs which make the shocked shells optically thick to Thomson scattering,
if  $\eta$ is below a certain value for which the comoving radiation
compactness parameter $\ell' \sim (L\st/m_e c^3 r) \simg 1$ (\Mesz \&
Rees 2000). Earlier simulations involving randomly ejected shells and
(baryonic) electron scattering in shocks have indicated a tendency
for more variable light curves arising in more distant shocks
(Panaitescu \etal  2000; Spada \etal 2000;
Ramirez-Ruiz \& Lloyd-Ronning 2002), as expected since closer in the
scattering depth is larger. Similar results are obtained numerically
when pair formation is included, e.g. Kobayashi \etal (2002). Here we
pursue a simplified analytical description. For shocks at increasing
radii, the pair comoving scattering depth of the shells eventually drops
to unity, $\tau_\pm'\sim
n_\pm' \st (r/\Gamma_c)$ $\sim (E_\pm \st / 4\pi r^2 \Gamma_c)\sim 1$
at a characteristic limiting pair-producing shock radius
\beq
\rpm \sim (E_\pm \st/ 4\pi  m_e c^2 \Gamma_c)^{1/2} \sim
     3\times 10^{14}(\epse3 \epsi4
N_2^{-1}\E54)^{1/2}(\GM3\Gm2)^{-1/4}~\cm ,
\label{eq:rpm}
\enq
(Kobayashi \etal 2002) where $\Gamma_m=10^2 \Gm2$, $\Gamma_M=10^3 \GM3$.
This is in the discrete shell regime, as opposed to the wind regime
used \Mesz \& Rees 2000; in general the shell regime pair density exceeds the
wind regime pair density by a factor $(t_w/t_v N)(\Gamma_c/\Gamma_m)^2 \geq 1$,
as expected since the same kinetic energy density is concentrated in shells
rather than smoothed out.
At this radius both the comoving scattering time and the pair
formation time as well as the comoving pair annihilation time
$(n'_\pm \st c)^{-1}$ become equal to the comoving expansion time
$\rpm/c\Gamma_c$.

Shells  with Lorentz factors $\Gamma_M$ and $\Gamma_m$ ejected from a
starting radius $r_o$ at time intervals $t_v=\Delta_o/c$ collide at a
radius $r_{sh}\sim c t_v \Gamma_m^2 \simg r_o\eta^2$. If this shock
radius is outside the limiting pair-shock radius $\rpm$ given by equation
(\ref{eq:rpm}), pairs do not form in the shock, whereas in the opposite
case an optically thick pair region does form in the merged shell,
which expands until it reaches the radius $\rpm$.  The shocks which
occur outside the limiting pair-shock radius $\rpm$ are those for which
the corresponding shells started out from
$r_o$ with a minimum time difference $t_v > \tvpm$, where
\beq
\tvpm \sim \rpm/ c \Gamma_c^2 \sim
 0.2 (\epse3 \epsi4 N_2^{-1} \E54)^{1/2} \GM3^{-5/4}\Gm2^{-5/4}~\s.
\label{eq:tvpm}
\enq
If the shell ejection time differences $t_v$ have random realizations
between the minimum and maximum values $[t_{v,m},t_{v,M}]$ over the
total duration of the burst outflow $t_b \simg t_{v,M}$, out of the $N$
shells ejected there will be, on average, a fraction $(1-\tvpm/t_{v,M})$
which will lead to shocks outside the pair-shock radius. For a high
radiative efficiency, a fraction $0.5\eps_{\gamma,1/2}$ of which is
taken to be in the gamma-ray range, the isotropic-equivalent gamma-ray
fluence of the shocks above the limiting pair-shock radius is
approximately
\beq
\Egam \sim  (1/2) \eps_{\gamma,1/2}\eps_e \eps_i \Eiso
\left(1-{\tvpm \over t_{v,M}}\right) \sim 4\times 10^{52}
    \eps_{\gamma,1/2}\epse3 \epsi4 \E54
     \left(1-{\tvpm \over t_{v,M}}\right) {\rm ~erg}.
\label{eq:Egam}
\enq
Here $\tvpm/t_{v,M} \sim 2\times 10^{-2} (\epse3 \epsi4 N_2^{-1}
\E54 )^{1/2} t_{b,1}^{-1} \GM3^{-5/4}\Gm2^{-5/4} ~\siml 1$, with
$\tvpm \siml t_{v,M}\siml t_b$ where $t_b=10 t_{b,1}$ s is the burst
duration. In this simple model $\Egam$ represents the energy in the
variable $\gamma$-ray component of the burst, which arises above $\rpm$
and has variability on timescales $\simg \tvpm$. For $\tvpm \ll t_b$,
$\Egam$ is insensitive to $\Gamma_m$, but for short bursts or for
$\tvpm \simg 0.1 t_b$ there is a dependence of $\Egam$ on $\tvpm
\propto \Gamma_m^{-5/4}$. For small $\Gamma_m$ the typical pair-shock
radius $\rpm$ is further out, and the minimum variability timescale
$\tvpm$ is longer, with a consequently smaller variable $\Egam$ (fewer
shocks occur outside the more distant limiting pair-shock radius).
Larger $\Gamma_m$ lead to smaller limiting pair-shock radii,
shorter minimum variability timescales $\tvpm$ and larger isotropic
equivalent $\Egam$.

\section{Pair-producing Shocks and X-ray Rich Component}

For shocks occurring at $r_{sh}  \leq \rpm$, i.e. below the limiting
pair-shock radius where shocks can result in pair formation, the
scattering optical depth of the shocked shells can become $\tau'_\pm
\simg 1$ (even when the shock is above the baryonic photosphere given by
equations [\ref{eq:rphw},\ref{eq:rphsh}]). Pair formation causes the same
amount of shock energy to be spread among a larger number of particles
(new pairs) than in a purely baryonic outflow, and inverse Compton losses
due to up-scattering of its own photons (Ghisellini \& Celotti 1999)
become important. For a pair of shells undergoing a shock at $r<\rpm$,
it is expected that pair-production acts as a thermostat, and for comoving
compactness parameters $10 \siml \ell' \siml 10^3$ the comoving pair
temperature is $T'_\pm \sim 3-30$ keV, with $\tau'_\pm \sim$ few
(Svensson 1987). The scattering depth per shock due to pairs is
unlikely to be much
larger, because the scattering and the pair-formation cross sections are
comparable,  and unless dissipation and pair formation occurs uniformly
throughout the entire volume, down-scattering of photons above the
pair threshold rapidly leads to self-shielding (Ramirez-Ruiz \etal 2002).
As a specific example, we take $T'_\pm \sim 10$ keV and $\tau'_\pm \sim 3$
for one shock, producing a comoving spectrum peaked near $h\nu'\sim
3 k T'_\pm/{\tau'}_{\pm}^2 \sim 3 T'_{\pm,10} {\tau'}_{\pm,3}^{-2}$ keV.
Since at any time there may be more than one shock at $r<\rpm$, the
photons might encounter more than one shell before escaping (e.g. Spada
\etal, 2000), and would also undergo adiabatic cooling between the shells
by a factor $\sim r_{sh}/\rpm$. These two effects combined could lower the
escaping photon energy by a factor roughly estimated as $\zeta \sim 0.2
\zeta_{0.2}$. For a CM bulk Lorentz factor $\Gamma_c=
(\Gamma_M\Gamma_m)^{1/2} =300 \Gamma_{c,2.5}$, the observer-frame
pair-producing shock radiation peak is at
\beq
h\nu_{x,sh}\sim 100~ T'_{\pm,10} {\tau'}_{\pm,3}^{-2} \zeta_{0.2}
                             \Gamma_{c,2.5} [2/(1+z)]~\hbox{keV}~.
\label{eq:Exsh}
\enq
The peak energy (\ref{eq:Exsh}) is still substantially above the
black-body value $T_{sh,BB} \sim 4~(\epse3 \epsi4 \E54 N_2^{-1})^{-1/8}
\Gc25^{11/8}~{\rm keV}$. The BATSE distribution of peak energies (Preece
\etal 2000) has $\sim 10\%$ of bursts with $h\nu_{pk}\siml 100$ keV,
while the joint BATSE-BeppoSAX distribution of Kippen \etal (2002) shows
that most X-ray flashes (XRF) have peak energies in the 20-100 keV range,
with one exception at $3^{+4}_{-3}$ keV. A nominal value of $h\nu_{pk}
\sim 20$ keV can be obtained from equation (\ref{eq:Exsh}) with, e.g.
$\Gamma_c [2/(1+z)] \sim 60$

For completeness, we note that in the extreme case where pairs are
produced
uniformly throughout the entire volume, thermalization and an equilibrium
pair optical depth $\tau'_\pm \propto \ell'^{1/2}$ might be achieved,
where $\ell'$ is the comoving compactness (Guilbert, Fabian \& Rees
1984; Svensson 1987), although we expect $\tau_\pm$ in this case to be
much smaller.

The energy in the X-ray component from shocks arising below the limiting
pair-shock  radius $\rpm$, integrated over the burst duration $t_b$, is
the complement of the $\gamma$-ray energy produced in shocks arising above
$\rpm$ [c.f. equation (\ref{eq:Egam})]. The X-ray isotropic equivalent
fluence is
\beq
\Ex \sim (1/2)\eps_k\eps_e\eps_i \Eiso \left({\tvpm \over t_{v,M}}\right)
    \sim 4\times 10^{50} \eps_{k,1/2}(\epse3\epsi4\E54)^{3/2}
   N_2^{-1/2} t_{b,1}^{-1} \GM3^{-5/4}\Gm2^{-5/4} {\rm ~erg},
\label{eq:Ex}
\enq
where $\tvpm/t_{v,M}$ is given below equation (\ref{eq:Egam}),
$\eps_k=(1/2)\eps_{k,1/2}$ is an efficiency factor to account for
a fraction of order unity of the luminosity below $\rpm$ which is
re-converted into kinetic energy. This X-ray component could account
for most of the harder X-ray flashes (Heise \etal 2002; Kippen \etal
2002), with $h\nu_{pk} \simg 20$ keV, but if more XRFs are observed with
peaks as low as 3-5 keV this may require additional considerations.
On the other hand, the X-ray excess GRB  discussed by Preece \etal (1996)
have characteristics which, as a class, are close to those of the
$r_{sh} \siml \rpm$ pair-producing shocks discussed in this section.

The radiation from pair-shocks with $\tau_\pm \sim$ few would be subject
to a moderate amount of
time-smoothing $\Delta t_{var} \sim \Delta t_{var,orig} \tau_\pm$,
which partially degrades the original variability implied by the random
ejection  and shocking of shells. The smoothing would be more appreciable
at the shorter timescales, where it would lead to a filling in of the
narrow troughs between peaks (see also Panaitescu \etal 2000; Kobayashi
\etal 2002 and Ramirez-Ruiz \& Lloyd-Ronning 2002). This smoothing,
however, would not be expected to affect the coarser time structure of the
light curve, since not many scatterings are incurred before the photons
are advected with the flow.

\section{Variability Dependence on $\gamma$-ray Luminosity}

An observational correlation (Fenimore \& Ramirez-Ruiz 2001; Reichart,
\etal 2001) has been reported between the isotropic equivalent luminosity
$\Lgam$ and a variability measure $V$ of the $\gamma$-ray time profiles,
of the form
\beq
\Lgam \propto V^g, ~~~\hbox{where}~~~ g\simeq 3.3^{+2.5}_{-1.1}~.
\label{eq:LV}
\enq
The operational definition of $V$ is related to the normalized variance,
or the root mean square of the deviations from a smoothed light curve.
Observations of afterglows with breaks in the light curves are
believed to indicate the presence of a collimated jet-like outflow.
The simplest interpretation assumes a uniform jet cross-section
(independent of angle out to a jet edge $\theta$), in which case the
variety of break times indicates a variety of jet opening angles and the
data indicate an isotropic equivalent fluence anti-correlation with jet
opening angle $\theta$ (Frail \etal 2001), of the form $\Lgiso \propto
\theta^{-2}$.  Alternatively, the same data can be interpreted in
terms of a non-uniform (angle-dependent) cross section
jet with a universal jet pattern given by the same functional
relation between the energy output as a function of angle
$L_\gamma(\theta) \propto \theta^{-2}$ (Rossi \etal 2002; Zhang \&
\Mesz 2002; Salmonson \& Galama 2002). In the latter case the data is
interpreted as sampling different off-sets between the observer line of
sight and the jet axis. Norris (2002) and Salmonson \& Galama (2002)
analyzing the time-lag effects (see below) in a larger sample and
including redshift and luminosity function effects, argue for a somewhat
steeper angular index of -5/2, so
\beq
\Lgam \propto \theta^{-p}, ~~~\hbox{where}~~~ p\sim 2-2.5~~.
\label{eq:Ltheta}
\enq
In the previous sections we used an isotropic outflow but our results
continue to apply to the jet case as long as $\Gamma$ exceeds the inverse
of the jet opening angle $\theta$. The model interprets the variable
$\gamma$-ray luminosity as that portion which arise from shocks above
the limiting pair-shock radius $\rpm$, characterized by a minimum time
variability given by equation (\ref{eq:tvpm}),
$\tvpm \propto \Lgam^{1/2} \Gamma_m^{-5/4}$. This is based on
equation (\ref{eq:Egam}) relating $\Egiso \sim \Egam$ to $\E54$, and the
assumption that the average mean duration and redshift differences are
overshadowed by source-intrinsic variations in $E_\gamma$ and $\Gamma_m$,
so that approximately $\Lgam \propto \Egam$, and assuming that $\tvpm \ll
t_{v,M}$. The crucial dependence of $\tvpm$ is through $\Gamma_m$, rather
than $\Gamma_M$, since it is $\Gamma_m$ which determines the shock
radius.  It is reasonable to make the ansatz
$\Gamma_m \propto \theta^{-q}$, and a value $q\sim 2$ (e.g. MacFadyen
\& Woosley 1999; Kobayashi \etal 2002) follows from momentum conservation
in a ``sharp boundary jet" model where the energy and $\Gamma$ are
constant throughout its cross section but there is a range of opening
angles (e.g. Frail \etal 2001). In a ``universal jet profile" model where
$L$ and $\Gamma$ vary as function of $\theta$ (Rossi \etal 2002; Zhang \&
\Mesz 2002; Salmonson \& Galama 2002), a value $q\sim 2$ is also expected,
e.g. if the baryon loading in the jet are approximately independent of
$\theta$ but the energy varies as $\theta^{-2}$ (e.g equation [\ref{eq:Ltheta}]).
Setting $\Gamma_m\propto \theta^{-q}$, we have then $\tvpm \propto
\Lgam^{1/2}\Gamma_m^{-5/4}\propto \Lgam^{(2p-5q)/4p}$. The variability
$V$ of the gamma-ray light curves could be expected to scale, in an
approximate way, inversely proportional to a power of the minimum
variability timescale, $V\propto t_{v,min}^{-k}$. An approximate argument
shows that such an anticorrelation exists in the GRB data,
with an index $k\simeq 2/3 $ (e.g. Ioka \& Nakamura 2001; Plaga 2001).
Identifying $t_{v,min}$ with $\tvpm$, we have
\beq
\Lgam \propto V^{4p/[k(5q-2p)]}~.
\label{eq:Varlum}
\enq
If one takes $p=q$, which may be too idealized, the theoretical relation
is $\Lgam \propto V^2$, which is comparable to the lower limit fit of
Fenimore \& Ramirez-Ruiz (2001); the same result is obtained for
$p=5/2,~q=2,~k=1$.  Using the nominal values $p=5/2,~q=2,~k=2/3$ we get
$\Lgam \propto V^3$, in good agreement with the observed best-fit
relation $\Lgam \propto V^{3.3}$ of Fenimore \& Ramirez-Ruiz (2001).

\section{Discussion}
\label{sec:disc}

We have discussed the properties of the quasi-thermal baryonic
photospheric radiation component  in GRB. At high isotropic equivalent
luminosities, this component can dominate the non-thermal shock
component, and appears in the hard X-ray range in the source frame.
Such sources may be identified with the X-ray excess (Preece \etal 1996)
class of bursts. This photospheric quasi-thermal component can
inverse-Compton cool the non-thermal electrons in the shocks above it,
suppressing the MeV synchrotron component and enhancing an
inverse-Compton GeV non-thermal component. For high
dimensionless entropy $\eta=L/{\dot M} c^2$ and low ($z\siml 1$)
redshifts the quasi-thermal component appears at hard X-rays (and
in extreme cases at $\gamma$-rays), whereas for high redshifts (and/or
low $\eta$) it appears at soft X-rays.
We also have identified a new regime in the description of baryonic
photospheres from relativistic outflows, which is valid at moderate to
high $\eta$. The value of the final coasting Lorentz factor of the
outflow is not automatically the value it has when the flow becomes
optically thin, and has three different possible values
$\eta, ~\etast^2\eta^{-1/2}, ~\etast$, as discussed below equation
(\ref{eq:rph-eta}), depending on the value of the initial
dimensionless entropy $\eta$.

We have quantified the location of the outermost radius at which pairs
can form in internal shocks, and have argued that highly variable
gamma-ray light curves arise mostly from shocks above this limiting
pair-shock radius. The pair-shock radius determines the approximate ratio
of the fluences in a variable gamma-ray non-thermal component and in a
less variable softer ($\simg$ 20-25 keV) X-ray component. The latter
could also be responsible for X-ray excess GRB, and, for moderately low
bulk Lorentz factors or moderately high redshifts $\Gamma [2/(1+z)] \simg
60$, would be similar to most of the currently known X-ray flash (XRF)
bursts (Heise \etal 2001; Kippen \etal 2001), but additional
considerations may be needed to fit naturally the softest (3-5 keV) XRFs.
Smoother X-ray components are also obtained from closer-in shocks
neglecting pair formation (e.g. Ramirez-Ruiz \& Lloyd-Ronning 2002;
Spada \etal 2000), but smoothing and softening is stronger when there
is pair formation (see also Kobayashi \etal 2002).  This pair X-ray
component is generally softer than that of the baryonic photosphere.
When present, the pair photosphere enshrouds the baryonic photosphere,
but its modest opacity $\tau_\pm \siml 3$ is not sufficient to alter
significantly the spectrum of the baryonic photosphere. One or both of
these X-ray rich components may be present, depending on the bulk Lorentz
factor and isotropic equivalent total energy of the burst, and criteria
are discussed for the non-thermal $\gamma$-ray components to dominate over,
or be dominated by, these X-ray components.

An individual burst is characterized in Figure 1 by an average
$\eta= (L/{\dot M}c^2)$. The shock radius $r_{sh,o}$ plotted in Figure 1
is for the minimum variability time $t_v=t_o \sim 0.3$ ms, and a
second shock radius $r_{sh,3}$ is shown for  $t_v=10^3 t_o\sim 0.3$ s.
For $t_v\sim t_o$ and $\eta \siml \etat$ the corresponding shocks occur
below both the baryonic $r_{ph}$ and pair shock $r_{\pm}$ photospheres,
leading to X-ray rich bursts whose variability is partially suppressed.
For $t_v\simg 10^3 t_o$ and $\eta \simg \etat$ shocks occur at or above
both  $r_{ph}$ and $r_\pm$ leading to hard $\gamma$-rays with large
variabilty at $\simg 0.3$ s. An individual burst may have several
variability timescales present, leading to both types of components
simultaneously. Preponderance of one or the other leads to a short
timescale but low amplitude variability X-ray rich bursts or XRF, or to
a classical hard GRB with large amplitude variability mostly at
$\simg 0.3$ s. Roughly speaking, X-ray flashes would be expected from
the region $\eta <\eta_t \sim 250$, and classical GRB from $\eta> \eta_t$.
A baryonic photosphere component should be present at the begining of
bursts and X-ray flashes, and in the troughs between harder peaks due
to shock radiation.  However, the farther beyond the coasting radius
the photosphere occurs, the weaker its energy fraction is relative to
the shock and/or e$^\pm$ component, because its energy drops as $r^{-2/3}$.
Low values of $\eta$ lead to further-out, weaker baryon photospheres,
and at the same time to harder, relatively stronger shocks ocurring
closer in to the photosphere.  If long variability timescales are
absent and $\eta \ll \etat$ a soft baryon photosphere may be the most
prominent component, but its total energy would be very low.
A strong baryonic photosphere dominated burst (with quasi-thermal
$\simg$ MeV spectrum) is possible (for shock efficiencies $\siml 0.1$)
for $\eta \simg \etast^{4/3}$, and such a component may be detectable
already for $\eta \simg \etast \sim 10^3$. On the other hand, slower
$\eta \siml \etat \sim 250$ outflows are likelier to make X-ray rich
bursts through a pair-shock component.

We argue also that the relationship between variable gamma-ray radiation
and the limiting pair-shock radius leads, using the phenomenologically
inferred dependence  between isotropic luminosity and jet angle,
to a simple analytical interpretation for the observed
variability-luminosity relation $L_\gamma \propto V^3$ (e.g. Fenimore
and Ramirez-Ruiz 2001; see also Kobayashi \etal 2002).  The positive
correlation between variability and harder $\nu F_\nu$ peaks discussed
by Lloyd-Ronning \& Ramirez--Ruiz (2002) also finds a qualitatively
similar interpretation in terms of a higher variability corresponding
to closer-in shocks, which are more specifically in the present model
shocks occurring just above the limiting pair-forming shock radius.

Several physical explanations have been proposed for the presence of
a cutoff above about 1 Hz  in the power density spectrum of GRB light
curves (Beloborodov \etal 1998) in terms of baryonic electron scattering
(e.g. Panaitescu \etal 2000; Spada \etal 2000; Ramirez-Ruiz \&
Lloyd-Ronning 2002). Here we point out a different explanation for
this, which is based on the existence of a minimum $\gamma$-ray
variability timescale
(equation [\ref{eq:tvpm}]) above which shock radiation is free from
smoothing by opacity from pair-formation. This is seen in Figure 1, which
shows that shocks associated with the variability timescales $t_v \simg 1$ s
and $\eta\simg 150$ common among observed GRB occur above the pair photosphere.

\acknowledgements{This research is supported by NASA NAG5-9192, NAG5-9153
and the Royal Society. We are grateful to S. Kobayashi for valuable discussions.}

\bigskip
\begin{figure}[htb]
\centering
\epsfig{figure=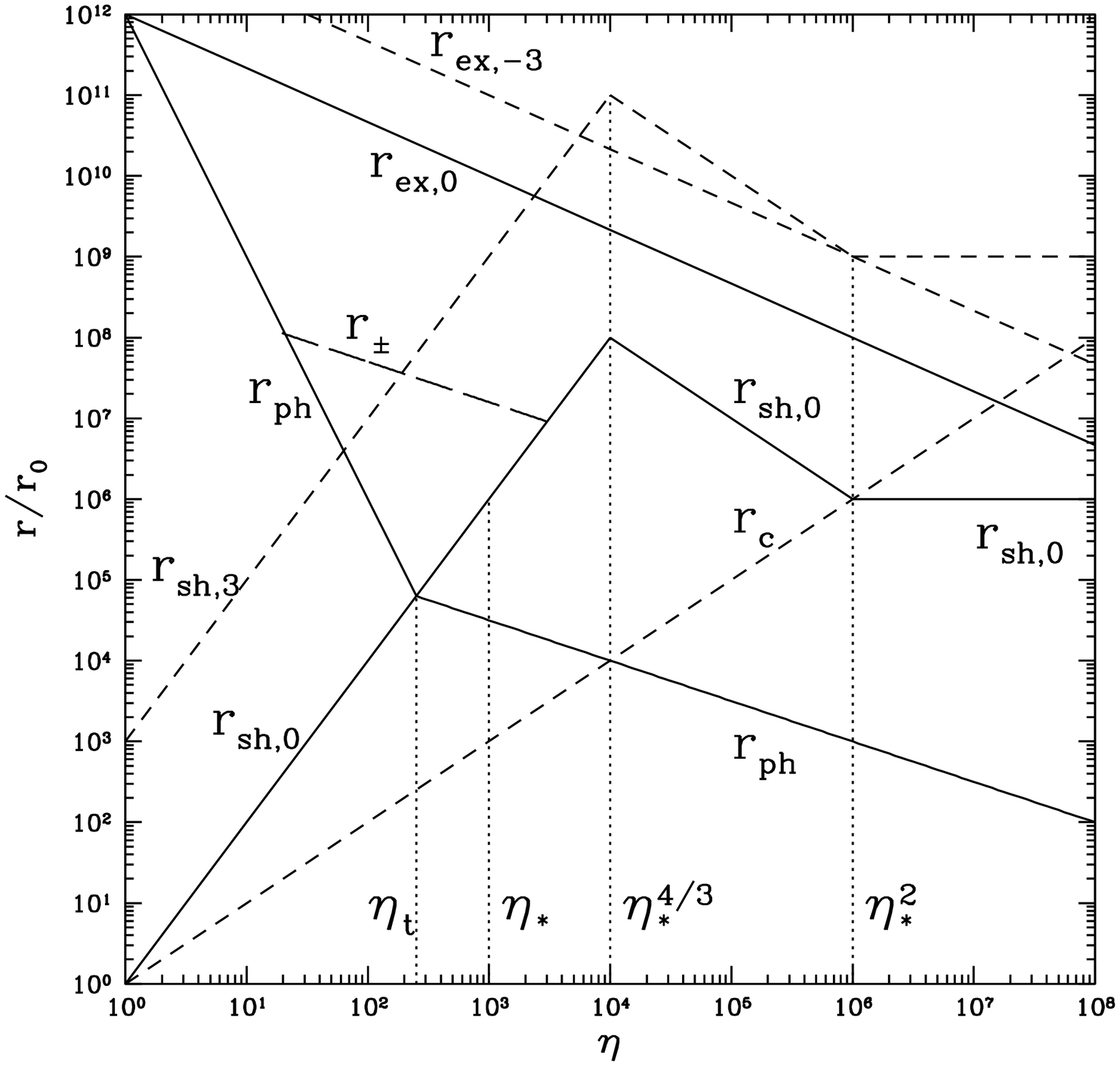,width=5.in,height=5.in}
\caption{
Schematic plot of the baryonic photospheric radius $r_{ph}$; the internal
shock radii $r_{sh,o}$, $r_{sh,3}$ for two different variability timescales
$t_o,~10^3 t_o$ where $t_o=0.3$ ms; the thin shell pair-producing photospheric
radius $r_\pm$; and the thin shell coasting radius $r_c$ starting at
$r_o=ct_o$, as a function of $\eta=L/{\dot M}c^2$. Distinguishing between
the continuous wind and discrete shell regimes leads to a new characteristic
value $\etat= \etast^{4/5}$ and a distinct regime between $\etast$ and
$\etast^{4/3}$ (equations (\ref{eq:etast},\ref{eq:etat}, see text).
Shocks can only occur above $r_{sh,o}$, and pair-forming shocks only occur
below $r_\pm$. Also shown is the external shock radius for two external
densities $n_{ext}=1,10^{-3}\cmcui$, giving an upper limit for internal
shock radii.
\label{fig:fig1}
}
\end{figure}

\end{document}